\documentclass[12pt]{article}
\usepackage[english]{babel}
\usepackage{amsfonts}
\usepackage{amsmath}
\usepackage{amssymb}
\usepackage{mathrsfs}
\usepackage{multicol}
\usepackage{fancybox}
\usepackage{dsfont}
\usepackage{color}
\usepackage{hyperref}
\usepackage{subfigure} 
\usepackage{graphicx} 
\usepackage{caption}
\usepackage{lineno}
\def\be{\begin{equation}}
\def\ee{\end{equation}}
\def\tl{\tilde}

\def\lm{\lambda}

\def\d'{``}
\def\i{\textrm{i}}
\def\be{\begin{equation}}
\def\ee{\end{equation}}
\def\bea{\begin{eqnarray}}
\def\eea{\end{eqnarray}}

\def\d'{``}
\begin{document}
%\linenumbers

\begin{center}
\Large{\bf{On an integrable discretisation of the Ablowitz-Ladik hierarchy}}
\end{center}

\begin{center}
{ {  Federico Zullo}}

{ School of Mathematics, Statistics and Actuarial Science,\\ University of Kent, CT2 7NF, Canterbury, Kent, UK.
\\~~E-mail: F.Zullo@kent.ac.uk}

\end{center}

\medskip
\medskip

\begin{abstract}
\noindent
Following the general results on the relationships about B\"acklund transformations (BTs) and exact discretisation given in a previous work \cite{FZBH}, we consider the Ablowitz-Ladik hierarchy and a corresponding family of BTs. After discussing the boundary conditions, we show how to get explicit transformations. The Hamiltonian properties of the maps and of the discrete flows are examined. The conditions on the parameters of the map giving exact discretisations are discussed. Finally, analytical and numerical examples are given.
\end{abstract}
%, with a proper choice of the parameters it is possible to obtain an

\bigskip\bigskip

\noindent


\section{Introduction} \label{intr}

There exist a fairly large number of papers explicitly or implicitly related to the subject of B\"acklund transformations (BTs) applied to the Ablowitz-Ladik (AL) model \cite{AL0}\cite{AL1}: here we just recall the paper by Suris \cite{S} and the more recent one by Tsuchida \cite{TT}. 

These two authors adopted the original approach of Ablowitz and Ladik \cite{AL2}, based on the zero-curvature representation of the model. More explicitly, suppose that our lattice model possess a Lax pair given by  $(L_k(\lm), M_k(\lm))$, where $k$ is the label for the lattice site and $\lm$ is the spectral parameter, i.e. an arbitrary constant independent of $k$ and $t$. Then, the linear problem associated to the model is given by \cite{AL1}:
\begin{equation}\label{LP}
L_k(\lm)\psi_k(\lm) =\psi_{k+1}(\lm), \qquad \frac{\partial \psi_k(\lm)}{\partial t}=M_k(\lm)\psi_k(\lm),
\end{equation} 

To the solution $\psi_k(\lm)$ of (\ref{LP}) we can associate another solution $\tl{\psi}_k$ of the same problem multiplying $\psi_k$ by a suitable matrix $D_k$, usually called Darboux or dressing matrix. The set of equations obtained \cite{AL2}, that is
\begin{equation}\label{DD}
L_k(\lm)\psi_k(\lm) =\psi_{k+1}(\lm), \qquad \tl{\psi}_k(\lm)=D_k(\lm)\psi_k(\lm),
\end{equation}
are the natural time discretisation of the system (\ref{LP}) if we consider the new solution $\tl{\psi}_k(\lm)$ as the old solution $\psi_k(\lm)$ but computed at the next time step. The compatibility condition implied by (\ref{DD}) is:
\begin{equation}\label{BTrelation}
\tl{L}_k(\lm) D_k(\lm) =D_{k+1}(\lm)L_k(\lm).
\end{equation}

The above procedure is broadly accepted to be a good starting point to obtain discretisations, but from this point on, there is a bifurcation in the literature results: there are various expressions for the dressing matrix $D_k$, in particular for its dependence on the spectral parameter, and the solutions of the \emph{overdetermined} system (\ref{BTrelation}) are presented in different forms. In \cite{S} Suris showed how the non local maps found by Ablowitz and Ladik \cite{AL2} can be factored in the composition of simpler and \emph{local} transformations, although defined only \emph{implicitly}.
%giving also the expression of the interpolating Hamiltonians at first order in the time step parameter. 
Then Tsuchida, in his interesting paper \cite{TT}, considers the relation \ref{BTrelation} for a given lattice system, not necessarily given by the AL model. Also he considers the expression obtained taking the determinant on both sides of \ref{BTrelation}:
\begin{equation*}
\det(\tl{L}_k(\lm))\det(D_k(\lm)) =\det(D_{k+1}(\lm))\det(L_k(\lm)),
\end{equation*}
showing how this is sufficient, under certain conditions, to obtain local expressions for the maps representing the discretisation of the underlying system. The author is aware of the fact that the discretisation obtained with this method are indeed BTs. However, he seems to not recognize that the algebraic equations solved by certain particular combinations of the dynamical variables appearing in the matrix $D_k$ (called \d'auxiliary variables'' in \cite{TT}), can be found by evaluating the determinant of the dressing matrix at a certain constant value of $\lm$ and then setting the result equal to zero. 

The relevance of this fact comes from the observation that it allows to obtain \emph{explicit} transformations, as shown for example in \cite{KS},\cite{RZK},\cite{FZGC}\cite{FZBH}. This may imply an improvement in the computations of the discretisation scheme. 

Basically the maps of Ablowitz and Ladik are non-local and implicit, the maps of Suris and Tsuchida are local and implicit, here we show how to get explicit maps, in general non local. Further, we show how our formulation gives a remarkable Hamiltonian characterization of the discrete flow. Indeed, since the BTs usually involve the whole hierarchy of commuting flows, it is reasonable to ask which of the flows of the hierarchy we are really discretising. More importantly, the question could be about the possibility to arrange the scheme so to discretise only a particular flow, e.g. the one corresponding to the physical Hamiltonian of the model. 

We will try to answer these questions following the results of \cite{FZBH}, where has been shown that the BTs for an integrable systems are deeply connected to the Hamiltonian structure of the system itself. Indeed they represent the integral curves of another integrable non-autonomous system that shares the conserved quantities with the original model: this observation can be used to obtain an explicit formula for the Hamiltonian interpolating the discrete flow defined by the iterations of the BTs. Furthermore, for any fixed orbit, the discretisation lies on a suitable (linear) combination of the integrals of motion: by choosing in a proper way the parameters appearing in the transformations it is then possible to make the discrete flow and the continuous one parallel for \emph{all} time. 

The paper is organized as follows: in section \ref{sec1} we briefly recall the results from \cite{FZBH}; in section \ref{sec2} we describe how to properly define the overall Lax matrix under periodic boundary conditions or in the case of an infinite support; in section \ref{sec3} we present the core findings of the work, that is the formulae for the explicit discretisation of the model and finally, in section \ref{sec4}, we discuss some analytical and numerical experiments.

\section{Hamiltonian flow associated to BTs}\label{sec1}
First of all, let us collect some known facts about BTs. Suppose to have an integrable system with $N$ degrees of freedom and $N$ quantities $\{H_i\}_{i=1}^N$ conserved and in involution. Denoting by $(p_i,q_i)_{i=1}^N$ the canonical coordinates, the Poisson structure is given by:
$$
\{p_i,q_j\}=\delta_{ij},\qquad \{p_i,p_j\}=\{q_i,q_j\}=0.
$$

A diffeomorphism from the phase space to itself, say
\begin{equation}\label{backlund}
(p_i,q_i)\overset{BT}{\Longrightarrow} (\tl{p}_i,\tl{q}_i), \qquad \textrm{where} \qquad
\left\{\begin{aligned}
&\tl{p}_i=f_i(p,q), \\
&\tl{q}_i=g_i(p,q), 
\end{aligned}\right.
\end{equation}
is a BTs if the following two conditions are satisfied:
\begin{itemize}
\item The transformations are canonical:
$$
\{\tl{p}_i,\tl{q}_j\}=\delta_{ij},\qquad \{\tl{p}_i,\tl{p}_j\}=\{\tl{q}_i,\tl{q}_j\}=0.
$$
\item The functions of the phase space given by the conserved quantities are invariant under the action of the map:
$$
H_i(\tl{p},\tl{q})=H_i(p,q).
$$
\end{itemize}

To simplify the notation we will drop hereafter the subscripts in the independent variables as above.
If the transformations are parametric the previous definition remains the same. In the rest of this section we assume to have parametric transformations with a parameter $\mu$; further, we assume that the BTs are connected to the identity, so that $\left.(\tl{p}, \tl{q})\right|_{\mu=0}=(p,q)$. As shown in \cite{FZBH}, these assumptions are not restrictive. 

One of the most interesting features of parametric BTs is the \emph{spectrality property} \cite{KS}: it is deeply related to the Hamiltonian structure of the transformations themselves. Let us call the parameter of the transformations $\mu$. Since the maps are canonical for every value of the parameter $\mu$, there exists a generating function $F$ such that \cite{FM}:
\begin{equation*} \begin{aligned}
&p_i=\frac{\partial F(q,\tl{q},\mu)}{\partial q_i},\\
&\tl{p}_i=-\frac{\partial F(q,\tl{q},\mu)}{\partial \tl{q}_i}.
\end{aligned}
\end{equation*}    
The function $F$ is a function of the $2N$ variables $(q_i,\tl{q}_i)_{i=1}^N$ and of the parameter $\mu$. Let us call $G(q,\tl{q},\mu)$ the partial derivative of $F$ with respect to $\mu$. We say that the BTs possess the spectrality property if, expressing $G$ as a function of the variables $(p_i,q_i)_{i=1}^N$ through the relations (\ref{backlund}), that is
$$
\Phi(p,q,\mu)\doteq G(q,g(p,q),\mu),
$$
the function $\Phi$ commutes with all the conserved quantities of the system:
$$
\{\Phi(p,q,\mu),H_i(p,q)\}=0, \qquad i=1,...,N. 
$$

An immediate consequence of this property is that the BTs are the integral curves, whit parameter $\mu$, of the non autonomous Hamiltonian system governed by the function $\Phi(p,q,\mu)$ \cite{FZBH}, that is
\begin{equation}\label{x} \begin{aligned}
&\frac{\partial \tl{q}_i}{\partial \mu}=\frac{\partial \Phi(\tl{p},\tl{q},\mu)}{\partial \tl{p}_i},\\
&\frac{\partial \tl{p}_i}{\partial \mu}=-\frac{\partial \Phi(\tl{p},\tl{q},\mu)}{\partial \tl{q}_i}.
\end{aligned}
\end{equation} 

As a remark, we stress that $\Phi(p,q,\mu)$, as a function on the phase space, is a function of the conserved quantities only \cite{FZBH}: 
$$\Phi(p,q,\mu)\doteq\breve{\Phi}(H,\mu).$$
 
This result has important consequences as regards the computational application of the BTs. Indeed, denoting by $p_i^{(n)}$ and $q_i^{(n)}$ the $n$-th iteration of the maps, the discrete trajectories defined by the set $\{p_i^{(n)}, q_i^{(n)}\}_{n=1}^T$ lie, for every value of $T$, exactly on the integral curves of the autonomous system 
\begin{equation} \begin{aligned}
&\frac{\partial \tl{q}_i}{\partial t}=\frac{\partial \mathcal{H}(\tl{p},\tl{q},\mu)}{\partial \tl{p}_i},\\
&\frac{\partial \tl{p}_i}{\partial t}=-\frac{\partial\mathcal{H}(\tl{p},\tl{q},\mu)}{\partial \tl{q}_i},
\end{aligned}
\end{equation} 
where the interpolating Hamiltonian $\mathcal{H}$ \cite{FZBH} is given by 
\begin{equation}\label{intham}
\mathcal{H}(p,q,\mu)=\int_{0}^{\mu} \Phi(p,q,\lm) d\lm .
\end{equation}

The concrete possibility to use the BTs in numerical applications comes from the following observation: for any function $\mathfrak{F}(p,q)$ on the phase space we can write
\begin{equation}\label{linearcomb}
\dot{\mathfrak{F}}=\{\mathcal{H},\mathfrak{F}\}=\sum_k c_k\{H_k,\mathfrak{F}\}, \quad \textrm{where}\quad c_k=\frac{\partial}{\partial H_k}\int_{0}^{\mu} \breve{\Phi}(H,\lm) d\lm .
\end{equation}
The functions $c_k$ are constants of motion, so they are constants along the trajectories of the model. Having in mind a numerical experiment, we must fix the numerical values of the initial conditions and those of the parameters. Then, the above formula tell us that we can treat the parameters $c_k$ just as constants: for any fixed set of initial conditions, the discrete trajectories coincide with the continuous ones given by the Hamiltonian
$$
\sum_{k=1}^N c_k H_k,
$$
where the numbers $c_k$ must be calculated according to the formulae (\ref{linearcomb}). 

Suppose now to have a multi-parametric set of BTs, depending both on the parameter $\mu$ and on the set $\{\lm_k\}_{k=1}^s$, for some integer $s\geq 1$. The interpolating Hamiltonian (\ref{intham}), as well as the constants $c_k$, will depend on this set of parameters. These additional degrees of freedom can be used to force the values of the $c_k$ to be equal to some pre-assigned set. We will see how this method works in the next sections, where we will construct a two-parameter family of BTs for the AL hierarchy.

\section{The Lax matrix and a comment on the boundary conditions}\label{sec2}
The AL model is described by the following equations of motion \cite{AL1}:
\begin{equation}\label{model}\begin{aligned}
&\dot{q}_k=q_{k+1}+q_{k-1}-2q_k-q_kr_k(q_{k+1}+q_{k-1}), \\
&\dot{r}_k=-r_{k+1}-r_{k-1}+2r_k+q_kr_k(r_{k+1}+r_{k-1}),
\end{aligned}
\end{equation}  
where $q_k(t)$ and $r_k(t)$ are dynamical variables on a lattice. It is possible to consider both periodic boundary conditions, that is $q_k=q_{k+N}$ and $r_k=r_{k+N}$ for some $N\in \mathbb{Z}$, and an infinite chain. For the infinite lattice we suppose that the variables possess a finite $L^1$ norm:
$$
\sum_{k}|q_k|<\infty, \qquad \sum_{k}|r_k|<\infty
$$
%Note that, as noted in \cite{APT}, this last case is relevant for the soliton solutions of the model, because these solutions possess an infinite support.\\ 
Let us discuss the definition of the Lax matrix of the model separately.

\textbf{Periodic case}. In the periodic case, the whole hierarchy of the commuting conserved quantities can be found from the invariants of the Lax matrix $L(\lm)$. This matrix is defined by the product
\begin{equation}\label{Lk}
L(\lm)=\stackrel{\curvearrowleft}{\prod_{k=1}^N}L_k(\lm), \qquad \textrm{with} \qquad L_{k}(\lm) =  \left( \begin{array}{cc} \lm & q_k\\ r_k& \lambda^{-1}\end{array} 
\right). 
\end{equation}

The determinant of $L(\lm)$, given by $\prod_{k=1}^N (1-q_kr_k)$, is a conserved quantity. The other $N-1$ conserved quantities appear in the Laurent expansion of the trace of $L(\lm)$:
\begin{equation}\label{Trace}
\textrm{Tr}(L(\lm))=\sum_{i=0}^N H_i\lambda^{N-2i}\;, \qquad H_0=H_N=1.
\end{equation}

The involutivity of these conserved quantities follows From the $r(\lm,\eta)$-matrix structure satisfied by the Lax matrix:
\begin{equation}
\{L(\lm)\stackrel{\otimes}{,}L(\eta)\}=[r,L(\lm)\otimes L(\eta)],
\end{equation}
where $r(\lm,\eta)$ is defined by \cite{S}:
$$
r(\lm,\eta)\doteq\left( \begin{array}{cccc} \frac{1}{2}\frac{\eta^2+\lm^2}{\eta^2-\lm^2} & 0&0&0\\ 0&-\frac{1}{2}&\frac{\lm \eta}{\eta^2-\lm^2}&0 \\ 0&\frac{\lm \eta}{\eta^2-\lm^2}&\frac{1}{2}&0\\0&0&0& \frac{1}{2}\frac{\eta^2+\lm^2}{\eta^2-\lm^2}  \end{array} \right).
$$

The above relations are equivalent to the following Poisson brackets for the dynamical variables of the model:
\begin{equation}\label{PB}
\{q_k,r_j\}=(1-q_kr_k)\delta_{kj}, \qquad \{q_k,q_j\}=\{r_k,r_j\}=0.
\end{equation}

The flow \ref{model} corresponds to the Hamiltonian $h$ given by:
\begin{equation}\label{physflow}
h=-H_1-H_{N-1}-2\log(\det(L)),\quad \textrm{where}\quad H_1=\sum_{k=1}^N q_kr_{k-1}, \; H_{N-1}=\sum_{k=1}^N r_kq_{k-1}
\end{equation}

\textbf{Infinite chain}. In this case the Lax matrix is given by the formula:
\begin{equation}\label{mon}
L(\lm)=\lim_{\substack{n\to \infty\\m\to -\infty}}\stackrel{\curvearrowleft}{\prod_{k=m}^n}L_k(\lm),
\end{equation}
whenever the limits exist. 

It is important to notice that the equation (\ref{BTrelation}), defining the BTs, is homogeneous in $L_k$: if we multiply the matrices $L_k$ by any scalar number $a$, the BTs are not affected by this operation. Let us denote by $L_{i,j}(\lm)$, $i,j=1,2$, the elements of the Lax matrix (\ref{mon}). From the homogeneity of (\ref{BTrelation}) it follows that the BTs are homogeneous function of degree zero of the elements $L_{i,j}(\lm)$; for example for $\tl{q}_k$ we will see that
\begin{equation}\label{BTs1}\begin{aligned}
\tl{q}_n=\tl{q}_n(q, r, \varphi_k(L_{i,j}(\lm_k)),\lm_k),
\tl{r}_n=\tl{r}_n(q, r, \varphi_k(L_{i,j}(\lm_k)),\lm_k),
\end{aligned}\end{equation} 
where $\lm_k$ are a set of free parameters and $\varphi_k(L_{i,j}(\lm_k))$ stands for a set of functions of the matrix elements $L_{i,j}$ evaluated in $\lm=\lm_k$. The functions $\varphi_k(L_{i,j}(\lm_k))$ must be all homogeneous of degree zero in the variables $L_{i,j}(\lm_k)$, that is
$$
\varphi_k(a L_{i,j}(\lm_k))=\varphi_k(L_{i,j}(\lm_k)), \qquad \forall a\in \mathbb{C}. 
$$  

The Lax matrix (\ref{mon}) is well defined for $|\lm|=1$ if the sequences $\{r_k\}$ and $\{q_k\}$ have finite $\ell^1$ norm (see \ref{app1}). However, the functions $\varphi_k(L_{i,j}(\lm_k))$ are well defined even for $|\lm_k|\neq 1$ in all the points where they are continuous. To clarify this point let us consider the sequence of matrices $T_N$:
$$
T^N=\stackrel{\curvearrowleft}{\prod_{k=-N}^N}L_k(\lm).
$$  

The corresponding elements can be written as
\begin{equation}
\begin{aligned} T^N_{11}&=\sum_{k=-N}^N l_{11}^{(k)}\lm^{2k+1},\\  T^N_{12}&=\sum_{k=-N}^N l_{12}^{(k)}\lm^{2k},\end{aligned}
\qquad  \begin{aligned} T^N_{21}&=\sum_{k=-N}^N l_{21}^{(k)}\lm^{2k},\\  T^N_{22}&=\sum_{k=-N}^N l_{22}^{(k)}\lm^{2k-1},\end{aligned}
\end{equation}
where the coefficients $l_{i,j}^{(k)}$ are functions of the dynamical variables. Consider, for example for $|\lm|<1$, a homogeneous function $\varphi(T^N_{ij}(\lm))$ of degree zero. Due to the absolute convergence of $T^N_{ij}$ for $|\lm|=1$ and using the continuity of the function $\varphi$, we can write
$$
\lim_{N\to \infty}\varphi(T^N_{ij}(\lm))=\lim_{N\to \infty}\varphi(\lm^{2N+1}T^N_{ij}(\lm))=\varphi(\lim_{N\to \infty}\lm^{2N+1}T^N_{ij}(\lm))=\varphi(L_{ij}(\lm)).
$$
The convergence of $\varphi(T^N_{ij}(\lm))$ for $|\lm|>1$ can be proved similarly. In the case of an infinite chain, the BTs (\ref{BTs1}) are then well defined also for values of the parameters $\lm_k$ inside or outside the unit circle.

In the following we will freely talk about the matrix $L(\lm)$: the results will apply independently of the boundary conditions assumed.

\section{Discretisation of the Ablowitz-Ladik hierarchy}\label{sec3}
Looking at the $\lm$ dependence of $L_k(\lm)$ (\ref{Lk}), we seek a dressing matrix $D_k$ of the form:
\begin{equation}\label{Dk}
D_{k}(\lm) =  \left( \begin{array}{cc} w^{-1}\left(\lm +\lm^{-1}a_k\right) & b_k\\ c_k& w\left(\lambda^{-1}+\lm d_k\right)\end{array} 
\right). 
\end{equation}
Here $w$ is a constant, independent on the lattice label $k$, introduced for later convenience. 

As noticed in the introduction, to obtain explicit transformations it is important to have a dressing matrix that is singular when $\lm$ assumes certain constant values. We will take a matrix $D_k(\lm)$ possessing two degenerate values of $\lm$, say $\lm_1$ and $\lm_2$: these values will appear also as free parameters of the transformations. Note that the constraints $\det(D_k(\lm_1))=0$ and $\det(D_k(\lm_2))=0$ leave only two free variables among the four appearing in \ref{Dk}. We can parametrize $D_k$ as follows:
\begin{equation}\label{Dk1}
D_{k}(\lm) =  \left( \begin{array}{cc} \frac{\lm}{w} +\frac{\lm_1\lm_2(\alpha_k\lm_2-\beta_k\lm_1)}{w\lm(\lm_2\beta_k-\lm_1\alpha_k)} & \frac{\lm_1^2-\lm_2^2}{\lm_2\beta_k-\lm_1\alpha_k}\\ \frac{\alpha_k\beta_k(\lm_1^2-\lm_2^2)}{\lm_1\lm_2(\lm_2\beta_k-\lm_1\alpha_k)}& w\left(\frac{1}{\lm}+\frac{\lm(\alpha_k\lm_2-\beta_k\lm_1)}{\lm_1\lm_2(\lm_2\beta_k-\lm_1\alpha_k)}\right)\end{array} 
\right), 
\end{equation}
where $\alpha_k$ and $\beta_k$ are two new variables.% Now we can look at the relation (\ref{BTrelation}): it must hold for every value of $\lm$, in particular also for $\lm=\lm_1$ and $\lm=\lm_2$. 
The matrices $D_k(\lm_1)$ and $D_k(\lm_2)$ possess a kernel by construction, respectively given by
\begin{equation*}
|\Omega_k^1>= \left( \begin{array}{c} w\\ \alpha_k \end{array} \right), \qquad  |\Omega_k^2>= \left( \begin{array}{c} w\\ \beta_k \end{array} 
\right). 
\end{equation*}
By applying $|\Omega_k^1>$ to the relation (\ref{BTrelation}) evaluated at $\lm=\lm_1$ we obtain
\begin{equation*}
\tl{L}_k(\lm_1)D_k(\lm_1)|\Omega_k^1>=0=D_{k+1}(\lm_1)\left(L_{k}(\lm_1)|\Omega_k^1>\right).
\end{equation*}
Up to an overall factor, $|\Omega_{k}^1>$ is unique. Then, from the previous equation, we can write
$$
L_{k}(\lm_1)|\Omega_k^1>=g_k|\Omega_{k+1}^1>,
$$
for some function $g_k$. More explicitly, after eliminating $g_k$, we obtain the recursion
\begin{equation}\label{arec}
\alpha_{k+1}=\frac{w(\lm_1 w r_k+\alpha_k)}{\lm_1(\lm_1 w+q_k\alpha_k)}=\frac{w}{\lm_1 q_k}-\frac{w^2(1-q_k r_k)}{q_k(\lm_1w+q_k\alpha_k)}
\end{equation} 
Exactly for the same reason, the variables $\beta_k$ must obey the recursion:
\begin{equation}\label{brec}
\beta_{k+1}=\frac{w(\lm_2 w r_k+\beta_k)}{\lm_2(\lm_2 w+q_k\beta_k)}=\frac{w}{\lm_2 q_k}-\frac{w^2(1-q_k r_k)}{q_k(\lm_2w+q_k\beta_k)}
\end{equation} 
\textbf{Remark}. If we wish to construct a map preserving the conserved quantities, the boundary values for $\beta_k$ and $\alpha_k$ are \emph{not} arbitrary. 

Take for example the periodic case. The Lax matrix of the model is given by $L=L_N...L_1$. Imposing a periodicity also on the dressing matrix, so that $D_{N+k}=D_k$ for every $k$, form the relations (\ref{BTrelation}) it follows that
\begin{equation}\label{LDDL}
\tl{L}(\lm)D_1(\lm)=D_1(\lm)L(\lm).
\end{equation} 
Since the eigenvalues of $L(\lm)$ are time-independent, the isospectral equation (\ref{LDDL}) implies the preservation of the conserved quantities under the action of the map. The periodicity of $D_k$ means that the variables $\alpha_k$ and $\beta_k$ are periodic as well: the relations (\ref{arec}) and (\ref{brec}) are then periodic continued fraction. This means that the variables $\alpha_k$ and $\beta_k$ are solutions of quadratic equations.
To find out these quadratic equations, we must recall that $D_1(\lm_1)$ and $D_1(\lm_2)$ are degenerate matrices. Again, applying the kernels $|\Omega_1^1>$ and $|\Omega_1^2>$ to the equation (\ref{LDDL}) evaluated respectively at $\lm=\lm_1$ and $\lm=\lm_2$ we obtain
\begin{equation*}
\tl{L}(\lm_i)D_1(\lm_i)|\Omega_1^i>=0=D_{1}(\lm_i)\left(L(\lm_i)|\Omega_1^i>\right), \qquad i=1,2.
\end{equation*}
This time we conclude that $|\Omega_1^i>$ is an eigenvalue of $L(\lm_i)$, so we can write
$$
L(\lm_i)|\Omega_1^i>=\gamma_i|\Omega_1^i>, \qquad i=1,2,
$$
from which, eliminating $\gamma_i$, we obtain the quadratic equations solved by $\alpha_1$ and $\beta_1$
\begin{equation}\label{quaeq}\begin{split}
&L_{1,2}(\lm_1)\alpha_1^2+(L_{1,1}(\lm_1)-L_{2,2}(\lm_1))w\alpha_1-L_{2,1}(\lm_1)w^2=0,\\
&L_{1,2}(\lm_2)\beta_1^2+(L_{1,1}(\lm_2)-L_{2,2}(\lm_2))w\beta_1-L_{2,1}(\lm_2)w^2=0.
\end{split}
\end{equation}

Vice-versa, if we start from these boundary values we will obtain a periodic dressing matrix, since the values $\alpha_{N+k}$ and $\beta_{N+k}$ obtained from (\ref{arec}) and (\ref{brec}) are equal to those of $\alpha_{k}$ and $\beta_{k}$. Similarly, in the case of an infinite period, to take the limiting values of $\alpha_k$ and $\beta_k$, for $k\to -\infty$, as the solutions of the equations (\ref{quaeq}), implies the equivalence
$$
\lim_{k\to\infty}D_k(\lm)=\lim_{k\to-\infty}D_k(\lm),
$$  
which, in turns, implies the preservation of the conserved quantities under the action of the BTs. These results agree with the observations made by Tsuchida about the boundary conditions for dressing matrices of lattice systems \cite{TT}. 

From the expression (\ref{BTrelation}), using (\ref{brec}) and (\ref{arec}), we can write the explicit BTs as:
\begin{equation}\label{BTs}\begin{aligned}
&\tl{q}_k=\frac{\lm_1\lm_2\left(w(\lm_1^2-\lm_2^2)+q_k(\lm_1\alpha_k-\lm_2\beta_k)\right)}{w^2\left(\lm_1\beta_k-\lm_2\alpha_k\right)}\\
&\tl{r}_k=\frac{w(\lm_1^2-\lm_2^2)\alpha_k\beta_k+w^2r_k\lm_1\lm_2\left(\lm_1\alpha_k-\lm_2\beta_k\right)}{\lm_1^2\lm_2^2\left(\lm_1\beta_k-\lm_2\alpha_k\right)}
\end{aligned}\end{equation}
where $\alpha_k$ and $\beta_k$ are given by the expressions (\ref{arec}) and  (\ref{brec}), with the boundary conditions fixed by the quadratic equations (\ref{quaeq}). %Shortly we will see how to determine the sign in the solution of this quadratic equations. 
Before to discuss how to determine the sign in the solution of the quadratic equations (\ref{quaeq}), let us make a remark.

A natural question can arise if one consider other type of boundary values for the variables $\alpha_k$ and $\beta_k$: one expects that the transformations will me no more energy preserving; we will return on this point in section \ref{sec4}, showing indeed how in this case it is possible to obtain the one-soliton solution from the \d'vacuum'' solution ($q_k=r_k=0$ for every $k$). 

Now we are interested in the Hamiltonian flows generated by the transformations (\ref{BTs}), as outlined in section \ref{sec1}. We must check: 1) if the transformations are connected to the identity for some values of the parameters; 2) if the spectrality property holds \cite{FZBH}. Let us see the first point. We take opposite signs in the solutions of the quadratic equations (\ref{arec}-\ref{brec}) for $\alpha_1$ and $\beta_1$ (or $\alpha_{-\infty}$ and $\beta_{-\infty}$ in the case of an infinite lattice), that is:
\begin{equation}\label{albe}\begin{split}
&\alpha_{1}=w\left.\left(\frac{L_{22}-L_{11}+\Delta}{2L_{12}}\right)\right|_{\lm=\lm_1},\qquad \beta_{1}=w\left.\left(\frac{L_{22}-L_{11}-\Delta}{2L_{12}}\right)\right|_{\lm=\lm_2},\\ & \textrm{with} \qquad \Delta^2(\lm)\doteq \textrm{Tr}(L(\lm))^2-4\det(L(\lm)).
\end{split}\end{equation}
Also, let us pose:
\begin{equation}\label{l1l2}
\lm_1=e^{\mu+\i\theta},\qquad \lm_2=e^{-\mu+\i\theta},\qquad w=\i e^{\i\theta} 
\end{equation}
In the limit $\mu\to 0$, from (\ref{BTs}) we obtain $\tl{q}_k=q_k$, $\tl{r}_k=r_k$ for every $k$. So the transformations are connected to the identity.

Let us check the spectrality property. We need to find the generating function of the canonical transformations (\ref{BTs}). This is given by the function $F(r,\tl{r})$ satisfying the equation
\begin{equation}\label{gfun}
dF(r,\tl{r})=\sum_{k}\frac{\ln(1-\tl{q}_k\tl{r}_k)}{\tl{r}_k}d\tl{r}_k-\frac{\ln(1-q_kr_k)}{r_k}dr_k .
\end{equation} 
Indeed, directly from the Poisson structure (\ref{PB}), it is possible to check that $-\frac{\ln(1-q_nr_n)}{r_n}dr_n$ is a canonical one-form. From Appendix \ref{app2} the function $F(r,\tl{r})$ is given by
\begin{equation}\label{genfun} \begin{split}
F(r,\tl{r})&=\sum_n f_k(\tl{r}_k\lm_2^2)+f_k(\tl{r}_k\lm_1^2)+g_k(\tl{r}_k,r_{k+1}w^2)+g_k(\tl{r}_k\lm_1^1\lm_2^2,r_{k-1}w^2)+ \\
&-g_k(\tl{r}_k\tl{r}_{k+1}\lm_1^2\lm_2^2,r_k r_{k+1}w^4)+\ln(r_kw^2)\left(\ln(r_{k-1}w^2)+\frac{1}{2}\ln(r_{k}w^2)+\ln(r_{k+1}w^2)\right)+\\
&-\ln(\lm_1)^2-\ln(\lm_2)^2+\int^{r_kw^2}\frac{\ln(1-z)}{z}dz,
\end{split}\end{equation}
where we define the functions $f_k(a)$ and $g_k(a,b)$ by the relations
\begin{equation}
f_k(a)\doteq\int_{r_kw^2}^a \frac{\ln(z+r_kw^2)}{z},\qquad g_k(a,b)\doteq\int_{r_kw^2}^a \frac{\ln(z-b)}{z}.
\end{equation}
These integrals can be explicitly evaluated in term of logarithm and dilogarithm functions.

The variable canonically conjugate to the parameter $\mu$ is, by definition, the derivative of the generating function (\ref{genfun}) with respect to $\mu$. Inserting (\ref{l1l2}) in (\ref{genfun}) and taking the derivative, we obtain
\begin{equation}\label{imph}
\frac{\partial F}{\partial \mu}=2\sum_n \ln\left(\frac{\tl{r}_n e^{\mu}-r_n e^{-\mu}}{\tl{r}_n e^{-\mu}-r_n e^{\mu}}\right).
\end{equation}
This expression has to be evaluated explicitly in terms of the untilded variables $r_k$ and $q_k$. As shown in Appendix \ref{app2}, the result can be written as
\begin{equation}\label{spec}
\Phi_{AL}(r,q,\mu,\theta)\doteq \frac{\partial F}{\partial \mu}=2 \ln\left(\frac{\textrm{Tr}\left(L(e^{-\mu+\i\theta})\right)-\Delta(e^{-\mu+\i\theta})}{\textrm{Tr}\left(L(e^{\mu+\i\theta})\right)+\Delta(e^{\mu+\i\theta})}\right),
\end{equation}
where $\Delta(\lm)$ is defined in (\ref{albe}). It is now evident that indeed this expression depends on the dynamical variables only as combinations of conserved quantities, since the trace and the determinant of $L(\lm)$ are time-independent. From equation (\ref{spec}), comparing with (\ref{x}) and (\ref{intham}), we get the following two propositions:\\
\textbf{Proposition 1}. The BTs (\ref{BTs}) are the integral curves, with parameter $\mu$, of the non-autonomous Hamiltonian system of equations governed by the Hamiltonian $\Phi_{AL}(\tl{r},\tl{q},\mu,\theta)$:
\begin{equation} \begin{aligned}
&\frac{\partial \tl{r}_k}{\partial \mu}=(1-\tl{q}_k\tl{r}_k)\frac{\partial \Phi_{AL}(\tl{r},\tl{q},\mu,\theta)}{\partial \tl{q}_k},\\
&\frac{\partial \tl{q}_k}{\partial \mu}=-(1-\tl{q}_k\tl{r}_k)\frac{\partial \Phi_{AL}(\tl{r},\tl{q},\mu,\theta)}{\partial \tl{r}_k}.
\end{aligned}
\end{equation} 
The integral curves are identified by the initial values, given by $\tl{r}_k|_{\mu=0}=r_k$ and $\tl{q}_k|_{\mu=0}=q_k$.\\
\textbf{Proposition 2}. The discrete trajectories obtained by iterating the map (\ref{BTs}) lie, for every choice of initial values $(r_k,q_k)$ and of parameters $(\mu,\theta)$, on the respective trajectories of the following Hamiltonian system:
\begin{equation}\label{Hdiseq}
\begin{aligned}
&\frac{\partial r_k}{\partial t}=(1-q_k r_k)\frac{\partial \mathcal{H}}{\partial q_k},\\
&\frac{\partial q_k}{\partial t}=-(1-q_k r_k)\frac{\partial \mathcal{H}}{\partial r_k},
\end{aligned}\qquad \textrm{where}\qquad \mathcal{H}=\int_0^\mu\Phi_{AL}(r,q,\lm,\theta)d\lm .
\end{equation} 
Taking into account (\ref{spec}), after some manipulations the formula for the interpolating Hamiltonian $\mathcal{H}$ (\ref{Hdiseq}) can be also written as:
\begin{equation}\label{intham1}
\mathcal{H}=-2\int_{-\mu+\i\theta}^{\mu+\i\theta}\textrm{arccosh}\left(\frac{\textrm{Tr}(L(e^{\lm}))}{2\sqrt{\det(L(e^{\lm}))}}\right)d\lm.%\frac{d\lm}{\lm}
\end{equation} 

A last remark: it is possible to get the reduction $r_k=-q_k^{*}$ in the formulae (\ref{BTs}) for BTs, relevant for physical applications. In the previous formula and in the rest of the paper we use the asterisk to denote complex conjugation. It is easy to show that if $r_k=-q_k^{*}$, then the elements of the Lax matrix evaluated at $\lm=\lm_1=e^{\mu+\i\theta}$ and $\lm=\lm_2=e^{-\mu+\i\theta}$ obey the following relations:
\begin{equation*}
\begin{aligned}
&L_{1,1}(\lm_2)=L^{*}_{2,2}(\lm_1),\\
&L_{2,1}(\lm_2)=-L^{*}_{1,2}(\lm_1),
\end{aligned}\qquad \begin{aligned}
&L_{1,2}(\lm_2)=-L^{*}_{2,1}(\lm_1),\\
&L_{2,2}(\lm_2)=L^{*}_{1,1}(\lm_1).
\end{aligned} 
\end{equation*}
%$$
%L_{1,1}(\lm_2)=L^{*}_{2,2}(\lm_1),\quad L_{1,2}(\lm_2)=-L^{*}_{2,1}(\lm_1), \quad L_{2,1}(\lm_2)=-L^{*}_{1,2}(\lm_1), \quad L_{2,2}(\lm_2)=L^{*}_{1,1}(\lm_1).
%$$
These relations entails $\Delta^2(\lm_1)=\left(\Delta^2(\lm_2)\right)^{*}$ for the function $\Delta(\lm)$ defined in (\ref{albe}). The expressions for $\alpha_1$ and $\beta_1$, or $\alpha_{-\infty}$ and $\beta_{-\infty}$ in the case of an infinite lattice, are explicitly given by (\ref{albe}). By choosing $\Delta(\lm_2)=-\left(\Delta(\lm_1)\right)^{*}$ we obtain $\beta_1=-\frac{1}{\alpha_1^{*}}$, and, from the recursions (\ref{arec}-\ref{brec}), $\beta_k=-\frac{1}{\alpha_k^{*}}$ for every $k$. Inserting these last equations in (\ref{BTs}) one has $\tl{r}_k=-\tl{q}_k^{*}$ provided that $r_k=-q_k^{*}$.

Taking into account (\ref{l1l2}), the reduced BTs for the variables $q_k$ read:
\begin{equation}\label{rBTs}
\tl{q}_k=\frac{\left(\cosh(\mu)(|\alpha_k|^2+1)+\sinh(\mu)(|\alpha_k|^2-1)\right)q_k+2\i e^{2\i\theta}\sinh(2\mu)\alpha_k^{*}}{\cosh(\mu)(|\alpha_k|^2+1)-\sinh(\mu)(|\alpha_k|^2-1)}.
\end{equation}
where in the definition of the variable $\alpha_k$ (\ref{arec}) and its boundary value (\ref{albe}) the reductions $\beta_k=-\frac{1}{\alpha_k^{*}}$ and $r_k=-q_k^{*}$ must be taken into consideration.

\section{Numerical experiments and an example}\label{sec4}
As a first example, we take three interacting bodies. The trace of $L(\lm)$ (\ref{Trace}) is now explicitly given by:
$$
\textrm{Tr}(L(\lm))=\lm^3+(r_1q_2+r_2q_3+r_3q_1)\lm+\frac{q_1r_2+q_2r_3+q_3r_1}{\lm}+\frac{1}{\lm^3}.
$$ 
Taking into account also the expression for the determinant of $L(\lm)$, the three integrals of motion are given by:
\begin{equation*}
\begin{aligned}
&H_1=r_1q_2+r_2q_3+r_3q_1,\\
&H_2=q_1r_2+q_2r_3+q_3r_1,\\
&H_3=(1-r_1q_1)(1-r_2q_2)(1-r_3q_3).
\end{aligned}
\end{equation*} 

According to formulae (\ref{Hdiseq}) and (\ref{intham1}) the discrete flow defined by the BTs (\ref{BTs}) lies exactly on the continuous flow generated by the Hamiltonian $\mathcal{H}$ given by
\begin{equation}
\mathcal{H}=-2\int_{-\mu+\i\theta}^{\mu+\i\theta}\textrm{arccosh}\left(\frac{e^{3\lm}+H_1e^{\lm}+H_2e^{-\lm}+e^{-3\lm}}{2\sqrt{H_3}}\right)d\lm . %-2\int_{-\mu+\i\theta}^{\mu+\i\theta}\textrm{arccosh}\left(\frac{\cosh(3\lm)+(H_1+H_2)\cosh(\lm)+(H1-H_2)\sinh(\lm)}{\sqrt{H_3}}\right)d\lm %\frac{d\lm}{\lm}
\end{equation}

The three constants of motion $c_1$, $c_2$ and $c_3$, given by formula (\ref{linearcomb}), are explicitly defined by the expressions
\begin{equation}\label{Hdiseq1}
\begin{aligned}
&c_1=\frac{\partial \mathcal{H}}{\partial H_1}=-2\int_{-\mu+\i\theta}^{\mu+\i\theta}\frac{e^{\lm}}{\Delta(e^{\lm})}d\lm, \qquad c_2=\frac{\partial \mathcal{H}}{\partial H_2}=-2\int_{-\mu+\i\theta}^{\mu+\i\theta}\frac{e^{-\lm}}{\Delta(e^{\lm})}d\lm,\\
%&c_2=\frac{\partial \mathcal{H}}{\partial H_2}=-2\int_{-\mu+\i\theta}^{\mu+\i\theta}\frac{e^{-\lm}}{\Delta(e^{\lm})}d\lm\\
&c_3=H_3\frac{\partial \mathcal{H}}{\partial H_3}=\int_{-\mu+\i\theta}^{\mu+\i\theta}\frac{\textrm{Tr}(L(e^{\lm}))}{\Delta(e^{\lm})}d\lm ,
\end{aligned}
\end{equation}
where $\Delta(\lm)$ is defined in (\ref{albe}). Note that in the definition of $c_3$ there is an extra factor $H_3$ in front of the partial derivative term. In this way we take into account the fact that in the physical Hamiltonian $H_3$ appears as the argument of a logarithm (see eq. (\ref{physflow})). With this choice the continuous flow discretised is given by $c_1H_1+c_2H_2+c_3\ln(H_3)$.
   
\begin{figure}
\centering
\includegraphics[scale=0.7]{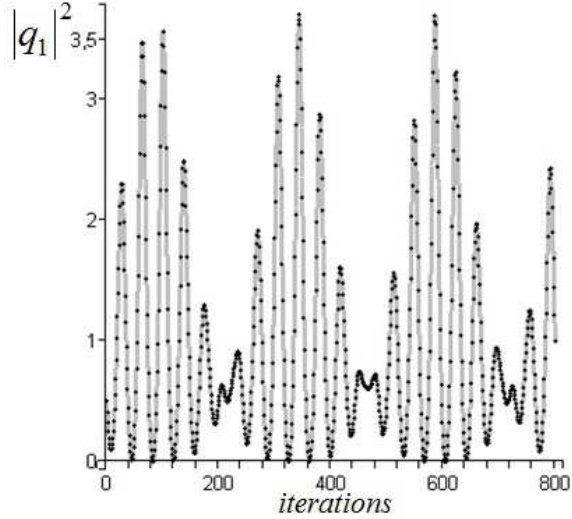}
\caption{\small{A continuous and discretised trajectory for $|q_1|^2$.}}
\label{fig1}
\end{figure}

The numerical values of the constants $c_1$, $c_2$ and $c_3$ are fixed by the values of the initial conditions and those of the parameters $\mu$ and $\theta$. In figure \ref{fig1} we plot the continuous trajectory of $|q_1|^2$ (in gray) and the iterations of the map (\ref{BTs}) (black dots), corresponding to the initial conditions $q_1=q_2=\frac{1}{\sqrt{2}}$, $q_3=\sqrt{\frac{7}{6}}$, and $r_k=-q_k^{*}$, $k=1,2,3$. In this case, the same plot could be obtained by iterating the map (\ref{rBTs}). The values of the conserved quantities are $H_1=H_2=-\frac{3}{2}$ and $H_3=6$; further, we chose the values $\mu=0.05$ and $\theta=0.1$, giving $c_1=0.00417-0.041645\i$, $c_2=-c_1^{*}$ and $c_3=-0.02234\i$. 

\begin{figure}
\centering
\includegraphics[scale=0.8]{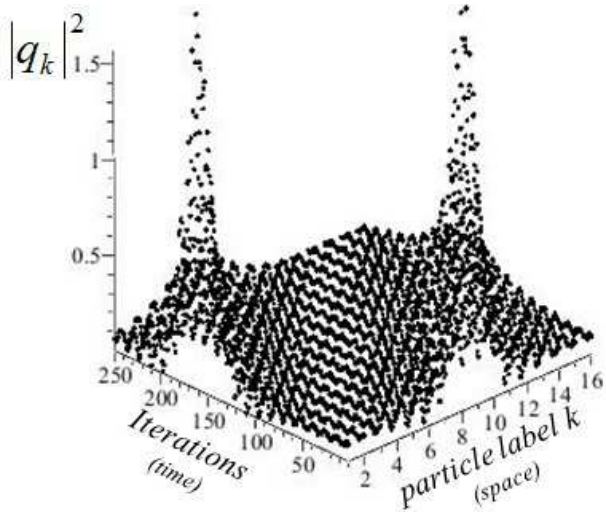}
\caption{\small{The set of trajectories for $|q_k|^2$, $k=1..16$.}}
\label{fig2}
\end{figure}

In the next example we take $N=16$. As initial conditions we chose $q_k=0.4+0.1\cos(-\pi+2\pi\frac{k-1}{15})$ and $r_k=-q_k^{*}$, corresponding, in the continuum limit, to $q(x,t=0)=0.4+0.1\cos(x)$ for $x\in (-\pi,\pi)$. The values of the parameters are $\mu=0.1$ and $\theta=0.02$. In figure \ref{fig2} we report the plots of $|q_k|^2$ for $k=1..16$, obtained from the map (\ref{rBTs}). A soliton structure emerges.

In this particular example we have sixteen conserved quantities and a BTs with two parameters. The constants of motion $c_k$, given by (\ref{linearcomb}), also depend on the parameters of the map. If we had sixteen parameters, by picking carefully their values it would be possible to obtain the particular combination of the constants $c_k$ corresponding to the physical flow (\ref{physflow}). A BTs with $2m$ parameters can be obtained by composing $m$ times the elementary BTs (\ref{rBTs}). However, the corresponding interpolating Hamiltonian will no longer be given by (\ref{intham1}). To generalize equation (\ref{intham1}) to the general case of BTs with $2m$ parameters appears to be difficult, but is an essential step to provide an exact discretisation of the physical flow (\ref{physflow}).

Our last example is about the \d'classical'' use of BTs. We want to obtain a non-trivial flow from the vacuum solution $q_k=r_k=0$. Considering the same boundary conditions on the variables $\alpha_k$ and $\beta_k$ as discussed in the previous section, that is $\alpha_{N+1}=\alpha_{1}$ and $\beta_{N+1}=\beta_{1}$ \emph{or} $\lim_{k\to\infty}\alpha_k=\lim_{k\to-\infty}\alpha_k$ and $\lim_{k\to\infty}\beta_k=\lim_{k\to-\infty}\beta_k$, we get only the trivial flow $\tl{q}_k=q_k=0$ and $\tl{r}_k=r_k=0$. This is clear from the fact that in this case the Hamiltonian function $\Phi$ (\ref{spec}) is independent of the dynamical variables, is just a constant. By a more physical point of view it is a consequence of the fact that the transformations, with the aforementioned boundary conditions for $\alpha_k$ and $\beta_k$, are energy conserving. To obtain a non-trivial flow from the vacuum solution we must relax these conditions. In the following, for simplicity, we will retrieve the original notation $\lm_1$ and $\lm_2$ for the parameters of the BTs (\ref{l1l2}). By taking $q_k=r_k=0$, the solutions of the recursions (\ref{arec}) and (\ref{brec}) are given by $\alpha_k=\alpha_0\lm_1^{-2k}$ and $\beta_k=\beta_0\lm_2^{-2k}$. In this example $\mu$ is no more a \d'time'' but just a parameter. Instead, the terms $\alpha_0$ and $\beta_0$ will contain the time dependences. To find out the expressions for $\alpha_0(t)$ and $\beta_0(t)$ we must look at the Poisson structure of the model. Just as an example we can take  
the well known semi-discrete version of the zero curvature condition corresponding to the equations of motion (\ref{model}) \cite{AL1},\cite{S}:
\begin{equation}
M_k=\i\left( \begin{array}{cc} 1-\lm^2+r_{k-1}q_k & \frac{q_{k-1}}{\lm}-q_k\lm\\ \frac{r_k}{\lm}-r_{k-1}\lm& \frac{1}{\lm^2}-1-r_{k}q_{k-1}\end{array} 
\right),
\end{equation}
where $M_k$ plays the role of the local Lax matrix associated to $L_k$ as in equation (\ref{LP}). Note that from the compatibility of equation (\ref{LP}) and equation (\ref{DD}) we can write:
\begin{equation}\label{Deq}
\dot{D}_k=\tl{M}_kD_k-D_kM_k
\end{equation}
If we evaluate (\ref{Deq}) in $\lm=\lm_1$, remembering that $D_k(\lm_1)|\Omega_k^1>=0$, we have
$$
\dot{D}_k|\Omega_k^1>=-D_k|\dot{\Omega}_k^1>=-D_kM_k|\Omega_k^1>\; \Longrightarrow D_{k}\left(|\dot{\Omega}_k^1>-M_k|\Omega_k^1>\right)=0.
$$
Since $\Omega_k^1$, the kernel of $D_k$, is uniquely defined up to a scalar, we obtain
$$
|\dot{\Omega}_k^1>=M_k|\Omega_k^1>+c|\Omega_k^1>,
$$
where $c$ is some scalar. From the previous equation we explicitly find
$$
c=\textrm{i}(\lm^2-1), \qquad   \alpha_0(t)=e^{\textrm{i}(\lm_1^2+\frac{1}{\lm_1^2}-2)t+a},
$$
where $a$ is a constant. Analogously, for $\beta_0(t)$ we have
$$
\beta_0(t)=e^{\textrm{i}(\lm_2^2+\frac{1}{\lm_2^2}-2)t+b}.
$$ 
Inserting these expressions in (\ref{BTs}) and taking into account the expressions (\ref{l1l2}),  we obtain for $\tl{q}_k$ and $\tl{r}_k$:
\begin{equation*}
\begin{aligned}
&\tl{q}_k=\frac{2\sinh(2\mu)e^{\textrm{i}\phi(n,t)}}{w\left(e^{b+\eta (n,t)}-e^{a-\eta (n,t)}\right)},\\
&\tl{r}_k=\frac{2w\sinh(2\mu)e^{a+b}e^{-\textrm{i}\phi(n,t)}}{\left(e^{b+\eta (n,t)}-e^{a-\eta (n,t)}\right)},
\end{aligned} \quad \textrm{with} \quad \left\{\begin{aligned} &\phi(n,t)\doteq(3+2n)\theta+2(1-\cosh(2\mu)\cos(2\theta))t,\\
& \eta (n,t)\doteq (2n+1)\mu+2t\sinh(2\mu)\sin(2\theta),    \end{aligned}\right.
\end{equation*}
i.e., a one soliton solution. It can be shown that this solution coincide with the simplest one found through the inverse scattering method (see e.g. \cite{AL1}, formulae 4.5-4.6).

\appendix
\section{Existence of the matrix $L(\lm)$ in the case of infinite support.}\label{app1}
To define the monodromy matrix for $|\lm|=1$ we can consider the recurrence $F_{n+1}=L_{n+1} F_n$, where the matrices $L_n$ are defined in (\ref{Lk}). To simplify the notation we will set 
\begin{equation*}
F_n(\lm)\doteq\left( \begin{array}{cc} A_n(\lm) & B_n(\lm)\\ C_n(\lm)& D_n(\lm)\end{array} 
\right). 
\end{equation*}   
The elements of the matrices $F_n(\lm)$ then satisfy:
\begin{equation}\label{lm1rec}
\begin{array}{c} A_{n+1}=\lm A_n+q_{n+1} C_n,\\ C_{n+1}=\frac{C_n}{\lm}+r_{n+1} A_n,\end{array} 
 \qquad  \begin{array}{c} D_{n+1}=\frac{D_n}{\lm}+r_{n+1} B_n,\\B_{n+1}=\lm B_n+q_{n+1} D_n .\end{array} 
\end{equation}

The previous recurrences are formally the same as those defining the Jost solutions in the inverse scattering method (see e.g. \cite{APT}, \cite{FT}): we can adopt the well known arguments on the existence of the Jost solution. In this appendix we will mainly follow \cite{APT}.

By denoting with $G^1$ the Green function for the recurrences solved by the sequences $A_n$ and $B_n$, and with $G^2$ the Green function for the recurrences solved by the sequences $C_n$ and $D_n$, we have   
\begin{equation}\label{gf}
\begin{array}{c} G^1_{n+1,k}=\lm G^1_{n,k}+\delta_{n,k},\\ G^2_{n+1,k}=\frac{G^2_{n,k}}{\lm}+\delta_{n,k},\end{array} 
\end{equation}
where $\delta_{n,k}$ is the usual Kronecker delta function. Passing to the Fourier space
\begin{equation*}
G^i_{n,k}=\frac{1}{2\pi\textrm{i}}\oint_{|w|=1}w^{n-1}\hat{G}^i_{w,k}dw, \qquad \delta_{n,k}=\frac{1}{2\pi\textrm{i}}\oint_{|w|=1}w^{n-k-1}dw .
\end{equation*}
and inserting the above formulae in (\ref{gf}) we find
\begin{equation*}
\hat{G}^1_{w,k}=\frac{1}{w^k(w-\lm)}, \qquad \hat{G}^2_{w,k}=\frac{1}{w^k(w-\lm^{-1})}.
\end{equation*}
Since $|\lm|=1$, we take a path avoiding the poles but enclosing them into the unit circle, obtaining:
\begin{equation}
G^1_{n,k}=\left\{ \begin{array}{c} \lm^{n-k-1}\quad k<n\\ 0\quad k\geq n\end{array} 
\right. ,\qquad G^2_{n,k}=\left\{ \begin{array}{c} \lm^{k-n+1}\quad k<n\\ 0\qquad\qquad k\geq n\end{array}
\right. ,
\end{equation}
so that we can write:
\begin{equation}\label{ACsol}
\begin{aligned} A_{n}&=f_0\lm^n+\sum_{k=-\infty}^{n-1}\lm^{n-k-1}q_{k+1} C_k,\\ C_{n}&=g_0\lm^{-n}+\sum_{k=-\infty}^{n-1}\lm^{k-n+1}r_{k+1} A_k ,\end{aligned} 
 \end{equation}
where $f_0$ and $g_0$ are two arbitrary constants. The solution to the equations \ref{ACsol} is formally given by two Neumann series:
\begin{equation}\label{Nse}
\begin{aligned} A_{n}=\sum_{p=0}^{\infty}f^p_n,\\  C_{n}=\sum_{p=0}^{\infty}g^p_n,\end{aligned} 
\quad \textrm{with}\quad \left\{ \begin{aligned} f_n^{p+1}&=\sum_{k=-\infty}^{n-1}\lm^{n-k-1} q_{k+1}g^p_k,\quad g^0_n=g_0\lm^{-n}. \\g_n^{p+1}&=\sum_{k=-\infty}^{n-1}\lm^{k-n+1}r_{k+1}f^p_k,\quad f_n^0=f_0\lm^n . \end{aligned} 
\right.
\end{equation}  
Using the result
$$
\sum_{k=-\infty}^{n}b_k\left(\sum_{j=-\infty}^{k}b_j\right)^m\leq \frac{1}{m+1}\left(\sum_{k=-\infty}^{n}b_k\right)^{m+1}
$$
valid for sequences $\{b_k\}$ with finite $\ell^1$ norm (see \cite{APT}, Lemma A.2), it is easy to recursively obtain the bounds:
\footnotesize{\begin{equation*}
\begin{aligned}|f_n^{2p}|&\leq |f_0|\frac{1}{p!}\left(\sum_{k=-\infty}^{n}|q_k|\right)^p \frac{1}{p!}\left(\sum_{k=-\infty}^{n}|r_k|\right)^p, \\ |g_n^{2p}|&\leq |g_0| \frac{1}{p!}\left(\sum_{k=-\infty}^{n}|q_k|\right)^p \frac{1}{p!}\left(\sum_{k=-\infty}^{n}|r_k|\right)^p,
\end{aligned} \quad \begin{aligned} |f_n^{2p+1}|&\leq |g_0|\frac{1}{(p+1)!}\left(\sum_{k=-\infty}^{n}|q_k|\right)^{p+1} \frac{1}{p!}\left(\sum_{k=-\infty}^{n}|r_k|\right)^p ,  \\  |g_n^{2p+1}|&\leq |f_0|\frac{1}{(p+1)!}\left(\sum_{k=-\infty}^{n}|r_k|\right)^{p+1} \frac{1}{p!}\left(\sum_{k=-\infty}^{n}|q_k|\right)^p.
\end{aligned}    
\end{equation*}}
\normalsize From the previous expressions it follows that the Neumann series (\ref{Nse}) converge for $|\lm|=1$ also in the limit $n\to \infty$. In exactly the same manner we can obtain the solution for the recursions on the right side of (\ref{lm1rec}).

\section{The generating function formula}\label{app2}
From the formula (\ref{gfun}) it follows that the generating function $F(r,\tl{r})$ of the canonical transformations (\ref{BTs}), solves the system
\begin{equation}\label{sys}\begin{aligned}
&r_k\frac{\partial F}{\partial r_k}=-\ln (1-q_k r_k),\\
&\tl{r}_k\frac{\partial F}{\partial \tl{r}_k}=\ln (1-\tl{q}_k\tl{r}_k).
\end{aligned}\end{equation} 

To express the right hand side of the previous equations as functions of $r$ and $\tl{r}$, we proceed in the following way. First, we solve the relation (\ref{BTrelation}) for the variables $q_k$, $\tl{q}_k$, $\beta_k$ and $\beta_{k+1}$. All these four variables will be functions of $\tl{r}_{k}$, $r_k$, $\alpha_k$, $\alpha_{k+1}$. We find the relations
\begin{equation}\label{qkqtk}\begin{aligned}
&q_k=\frac{w (\lm_1 w r_k+\alpha_{k}-\lm_1^2\alpha_k)}{\lm_1\alpha_k\alpha_{k+1}},\\
&\tl{q}_k=\frac{\lm_1 (\lm_1\lm_2^2 \tl{r}_k+w\lm_2^2\alpha_{k+1}-w\alpha_k)}{w^2\alpha_k\alpha_{k+1}},
\end{aligned}  
\qquad 
\begin{aligned}
&\beta_k =\frac{\lm_1^2\lm_2\alpha_k(w^2r_k+\lm_2^2\tl{r}_k)}{\lm_1\lm_2^2(w^2r_k+\lm_1^2\tl{r}_k)+w(\lm_2^2-\lm_1^2)\alpha_k},\\
&\beta_{k+1} =\frac{\lm_2\alpha_{k+1}(w^2 r_k+\lm_1^2\tl{r}_k)}{\lm_1(w^2r_k+\lm_2^2\tl{r}_k)+w(\lm_2^2-\lm_1^2)\alpha_{k+1}}.
\end{aligned}
\end{equation}
Substituting the index $k$ with the index $k+1$ in the expression for $\beta_k$ and comparing with the equivalent expression for $\beta_{k+1}$, we obtain the following expression for $\alpha_{k+1}$:
\begin{equation}\label{ak1}
\alpha_{k+1}=\frac{\lm_1\left(w^4r_k r_{k+1}-\lm_1^2\lm_2^2\tl{r}_k \tl{r}_{k+1}\right)}{w\left(\lm_1^2(\lm_2^2\tl{r}_{k+1}-\tl{r}_k)+w^2(\lm_1^2r_{k+1}-r_k)\right)}.
\end{equation}
The corresponding expressions for $\alpha_k$ can be found by substituting the index $k$ with $k-1$ in (\ref{ak1}). Inserting the formulae for $\alpha_k$ and $\alpha_{k+1}$ in the expressions (\ref{qkqtk}), we get $q_k$ and $\tl{q}_k$ explicitly in terms of the variables $r$ and $\tl{r}$. Collecting all together, the system (\ref{sys}) reads:
\begin{equation}\begin{aligned}
&r_k\frac{\partial F}{\partial r_k}=-\ln \frac{(w^2r_k+\lm_1^2\tl{r}_k)(w^2r_k+\lm_2^2\tl{r}_k)(w^2r_{k+1}-\tl{r}_k)(w^2\tl{r}_{k-1}-\lm_1^2\lm_2^2\tl{r}_k)}{(\lm_1^2\lm_2^2\tl{r}_k\tl{r}_{k-1}-w^4r_kr_{k-1})(\lm_1^2\lm_2^2\tl{r}_k\tl{r}_{k+1}-w^4r_kr_{k+1})}\\
&\tl{r}_k\frac{\partial F}{\partial \tl{r}_k}=\ln \frac{(w^2r_k+\lm_1^2\tl{r}_k)(w^2r_k+\lm_2^2\tl{r}_k)(w^2r_{k}-\tl{r}_{k-1})(w^2\tl{r}_{k}-\lm_1^2\lm_2^2\tl{r}_{k+1})}{(\lm_1^2\lm_2^2\tl{r}_k\tl{r}_{k-1}-w^4r_kr_{k-1})(\lm_1^2\lm_2^2\tl{r}_k\tl{r}_{k+1}-w^4r_kr_{k+1})}
\end{aligned}\end{equation} 
The formula (\ref{genfun}) for the generating function $F$ can be checked with a direct calculation.

\vspace*{6mm}
\noindent\textbf{Acknowledgments}

\noindent
I wish to acknowledge the financial support of the \d'Istituto Nazionale di Alta Matematica'' (Italian National Institute for High Mathematics) as an INdAM-COFUND Marie Curie Fellow.


\begin{thebibliography}{40}
\bibitem{AL0} Ablowitz M.J., Ladik J.F.: Nonlinear differential-difference equations, {\it J. Math. Phys.}, {\bf 16}, 598-603, 1975.
\bibitem{AL1} Ablowitz M.J., Ladik J.F.: Nonlinear differential-difference equations and Fourier analysis, {\it J. Math. Phys.}, {\bf 17}, 1011-1018, 1976.
\bibitem{AL2} Ablowitz M.J., Ladik J.F.: A nonlinear difference scheme and inverse scattering {\it Stud. Appl. Math.},
{\bf 55}, 213–29, 1976\\
 Ablowitz M.J., Ladik J.F.: On the solution of a class of nonlinear partial difference equations {\it Stud. Appl. Math.}, {\bf 57} 1–12 1977.
\bibitem{APT} Ablowitz M.J., Prinari B., Trubatch A.D.: Discrete and Continuous Nonlinear Schr\"odinger Systems, London Mathematical Society Lecture Note Series, Cambridge University Press, Cambridge, 2004. 
\bibitem{FM} A. Fasano, S. Marmi: Analytical Mechanics, Oxford University Press, New York, 2006. 
\bibitem{FT}
Faddeev L.D., Takhtajan L.A. 1987 Hamiltonian methods in the theory of
solitons,  Springer-Verlag.
\bibitem{KS} V.B. Kuznetsov, E.K. Sklyanin, On  B\"acklund Transformations for
  many-body systems, {\it J. phys. A: Math. Gen.}, {\bf 31}, 2241-2251, 1998. 
\bibitem{RZK} O. Ragnisco, F. Zullo: B\"acklund Transformation for the Kirchhoff Top” {\it SIGMA}, {\bf 7}, 001, 13 pages, 2011.
\bibitem{S} Suris Y.B.: A note on an integrable discretization of the nonlinear Schr\"odinger equation, {\it Inverse Problems}, {\bf 13}, 1121-1136, 1997.
\bibitem{TT} Tsuchida T.: A systematic method for constructing time discretizations of integrable lattice systems: \emph{local} equations of motion, {\it J. Phys. A: Math. Theor.}, {\bf 43}, 415202, 22 pages, 2010.
\bibitem{FZGC} Zullo F.: B\"acklund transformations for the elliptic Gaudin model and a Clebsch system, {\it J. Math. Phys.} {\bf 52}, 073507 2011.
\bibitem{FZBH} Zullo F.: B\"acklund transformations and Hamiltonian flows, to appear in {\it J. Math. Phys.}.  Preprint at arXiv:1207.0387.  
\end{thebibliography}
\end{document}